\documentclass[conference]{IEEEtran}
\IEEEoverridecommandlockouts
\usepackage{booktabs}

\usepackage{cite}
\usepackage{amsmath,amssymb,amsfonts}
\usepackage{algorithmic}
\usepackage{graphicx}
\usepackage{textcomp}
\usepackage[table]{xcolor}
\usepackage{multirow}
\usepackage{enumitem} 
\usepackage{microtype} 
\usepackage{stfloats}
\usepackage{float}
\usepackage{caption}
\begin{document}

\title{TriPAH: Imbalance-Aware Tri-Prompt Affinity Hashing for Cross-Modal Medical Retrieval%
\thanks{This work was supported in part by the National Natural Science Foundation of China under Grant 62536009, in part by the Hunan Provincial Graduate Research and Innovation Project under Grant CX20250248, and in part by the Henan Province Science and Technology Breakthrough Project under Grant 262102211053.}}

\author{
\IEEEauthorblockN{
Jiaming Bian,
Songming Li,
Yurui Song,
Yunfei Chen\IEEEauthorrefmark{1},
Yichao Cao\IEEEauthorrefmark{1},
Jun Long\IEEEauthorrefmark{1}
}
\IEEEauthorblockA{
Big Data Institute, Central South University, Changsha, China \\
\{bianjiaming, lisongming, csusyr\}@csu.edu.cn, yunfeichen.csu@gmail.com, caoyichao@csu.edu.cn, jlong@csu.edu.cn
}
\IEEEauthorblockA{\IEEEauthorrefmark{1}Corresponding authors}
}

\maketitle

\begin{abstract}
In the era of big medical data, efficient cross-modal retrieval is pivotal for evidence-based diagnosis and large-scale case management. Cross-modal medical hashing retrieval aims to enable efficient image–text search and support downstream tasks such as case-based reasoning and decision support by learning compact, semantically aligned binary codes. However, current methods suffer from semantic fragmentation due to noisy clinical language, long-tailed labels, and brittle quantization that weakens alignment. We propose TriPAH, a Tri-Prompt Affinity Hashing framework. TriPAH synthesizes ontology-grounded, patient-level prompts conditioned on normalized clinical cues to yield low-noise textual representations for initial alignment. A lightweight prompt–token mixer performs hierarchical, multi-granularity alignment and produces quantization-ready features under an asymmetric multi-task objective coupling multi-positive contrastive alignment, imbalance-aware classification, and progressive quantization regularization. A patient-level consistency module further stabilizes codes across complementary views. Extensive experiments on three public datasets demonstrate that TriPAH significantly outperforms state-of-the-art methods.
\end{abstract}


\section{Introduction}
\label{sec:intro}

Cross-modal medical retrieval aligns images and reports in a unified semantic space to support efficient bidirectional search for case evidence and diagnostic cues under practical computation and latency constraints \cite{Zhu2024TKDE_MMH_Survey}. Such methods map high-dimensional images and unstructured reports into compact metric-compatible representations for efficient cross-modal retrieval. In clinical scenarios, this capability is valuable because radiology reports constitute key diagnostic evidence and automated retrieval can accelerate evidence aggregation and review workflows \cite{Johnson2019SciData_MIMICCXR}.

Despite its promise, medical cross-modal retrieval remains challenging. Clinical reports often contain templated phrasing, ambiguity, and noise, while disease distributions are typically long-tailed and multi-label, with scarce annotations for rare classes \cite{Zheng2024NatComm_LongTailedDiagnosis}. In addition, supervision may vary across patient and exam levels, producing semantic gaps between modalities. These factors often lead to \textit{semantic fragmentation}, unstable similarity estimation, and degraded retrieval performance \cite{Wang2022EMNLP_MedCLIP}. Another key difficulty is \textit{directional asymmetry}: textual representations are usually more stable than visual ones, whereas images are more sensitive to acquisition conditions and phenotype variations. Under shared optimization, such asymmetry can amplify early quantization pressure and cause discretization collapse.

\begin{figure}[t]
  \centering
  \includegraphics[width=\linewidth]{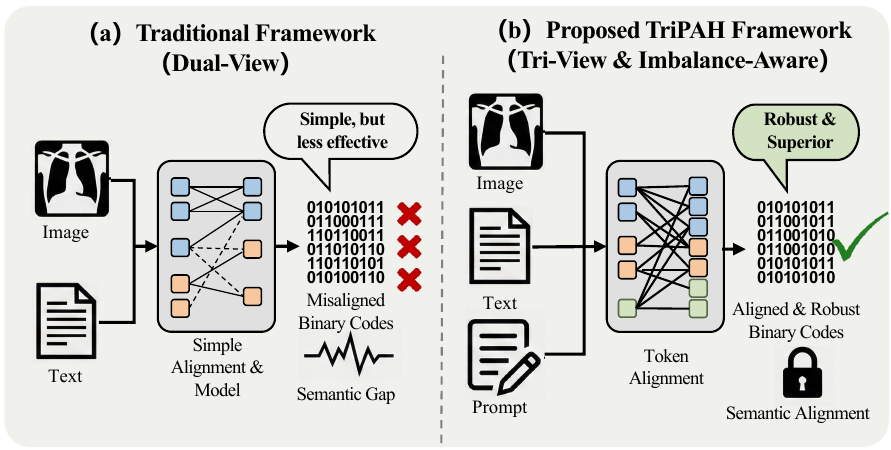}
  \vspace{-0.3cm}
  \caption{A conceptual comparison between traditional hashing frameworks and our proposed TriPAH framework.}
  \label{fig:comparison}
  \vspace{-0.5cm}
\end{figure}

Hash-based retrieval is particularly attractive for large-scale deployment because it learns compact binary codes for efficient Hamming search \cite{Cao2017ICCV_HashNet}. However, most existing frameworks rely on dual-branch alignment with shared quantization\cite{Jiang2017CVPR_DCMH}, which is less effective under noisy reports, long-tailed labels, and patient-level semantic variation. They lack an explicit corrective view for noisy text and rarely model imbalance or patient context, leaving directional gaps insufficiently addressed \cite{Zhu2024TKDE_MMH_Survey}.

To address these issues, we propose TriPAH, a Tri-Prompt Affinity Hashing framework for cross-modal medical retrieval. TriPAH introduces image, text, and learnable prompt views as complementary semantic anchors, and performs lightweight pairwise fusion before binary code generation. It further combines imbalance-aware multi-task optimization with branch-wise quantization annealing to improve robustness under long-tailed supervision and noisy clinical descriptions. In summary, the primary contributions of our work are as follows:

\begin{itemize}[itemsep=0pt, parsep=0pt]
  \item We propose a novel tri-view semantic fusion framework, which jointly models image features, text features, and a learnable prompt, effectively mitigating semantic fragmentation and stabilizing alignment for I$\rightarrow$T.
  \item We design an imbalance-aware multi-task hashing strategy that preserves class separability under long-tailed, multi-label regimes. This reduces early discretization collapse and alleviates rare-disease retrieval bias.
  \item We reshape instance descriptions using stochastic gating to curb bias through a patient context-adaptive prompt mechanism. This strengthens robustness and narrows the gap between bidirectional retrieval directions.
\item We conduct extensive experiments on ODIR-5K, MIMIC-CXR and IU-Xray datasets. Our results show consistent and substantial improvements of +26.2\%, +12.6\%, and +14.5\% in mean mAP.
\end{itemize}

\section{Related Work}

\subsection{Cross-modal Hashing \& Retrieval}
Multi-modal hashing balances representational capacity with retrieval efficiency. Recent cross-modal hashing methods have explored Transformer-based differentiable hashing and interactive fusion, such as DCHMT \cite{Tu2022ACMMM_DCHMT} and MITH \cite{Liu2023ACMMM_MITH}, as well as semantic-aware or contrastive optimization frameworks, such as DSPH \cite{Huo2024TCSVT_DSPH}, CMCL \cite{Wu2024TKDE_CMCL}, and CICH \cite{Luo2024TKDE_CICH}. In medical scenarios, MSACH \cite{Liu2025EAAI_MSACH} and MCPH \cite{Liu2024MICCAI_MCPH} begin to address noisy reports and modality gaps. However, most existing methods still insufficiently model long-tailed disease distributions and patient-level context, which limits robustness in clinical retrieval.

\subsection{Noisy Correspondence Learning}
Noisy correspondence degrades cross-modal representation quality. Standard solutions address this issue through meta-similarity correction or uncertainty-guided learning \cite{Han2023CVPR_MSCN,Zha2024SIGIR_UGNCL}. Unlike methods relying mainly on relabeling or sample refinement, medical-specific works such as MCPH \cite{Liu2024MICCAI_MCPH} improve robustness by combining prompt learning with noise-resistant constraints. Our method addresses noisy and long-tailed conditions through tri-view fusion and imbalance-aware optimization.

\subsection{Prompt Learning with VLMs}
Multi-modal prompt learning efficiently adapts pre-trained vision-language models to downstream tasks by injecting lightweight learnable parameters while keeping the backbone largely frozen \cite{Khattak2023CVPR_MaPLe}. In biomedicine, recent studies leverage domain-specific pre-training such as BiomedCLIP or task-aware prompting strategies for medical retrieval and hashing \cite{Zhang2023arXiv_BiomedCLIP,Zou2025CVPR_PromptHash}. However, integrating prompting with long-tail imbalance, noisy correspondence, and patient-level context remains underexplored. We address this gap through tri-view prompt fusion and quantization annealing.

\section{Method}

\subsection{Overview}
The goal of TriPAH is to map medical images and clinical reports into a unified, compact binary Hamming space while addressing three critical challenges: semantic fragmentation caused by noisy texts, misalignment due to lack of context, and retrieval bias from long-tailed disease distributions.
Let $\mathcal{O} = \{v_i, t_i, y_i\}_{i=1}^N$ denote a medical dataset with $N$ triplets, where $v_i$ represents the diagnostic image, $t_i$ is the clinical report, and $y_i \in \{0, 1\}^C$ is the multi-label annotation over $C$ disease categories.
Our objective is to learn two modality-specific hash functions, $H^v(v_i)$ and $H^t(t_i)$, which project inputs into $K$-bit binary codes $\mathbf{b}_i^v, \mathbf{b}_i^t \in \{-1, 1\}^K$, such that the Hamming distance effectively reflects the semantic similarity.

\begin{figure*}[t]
  \centering
  \includegraphics[width=\textwidth]{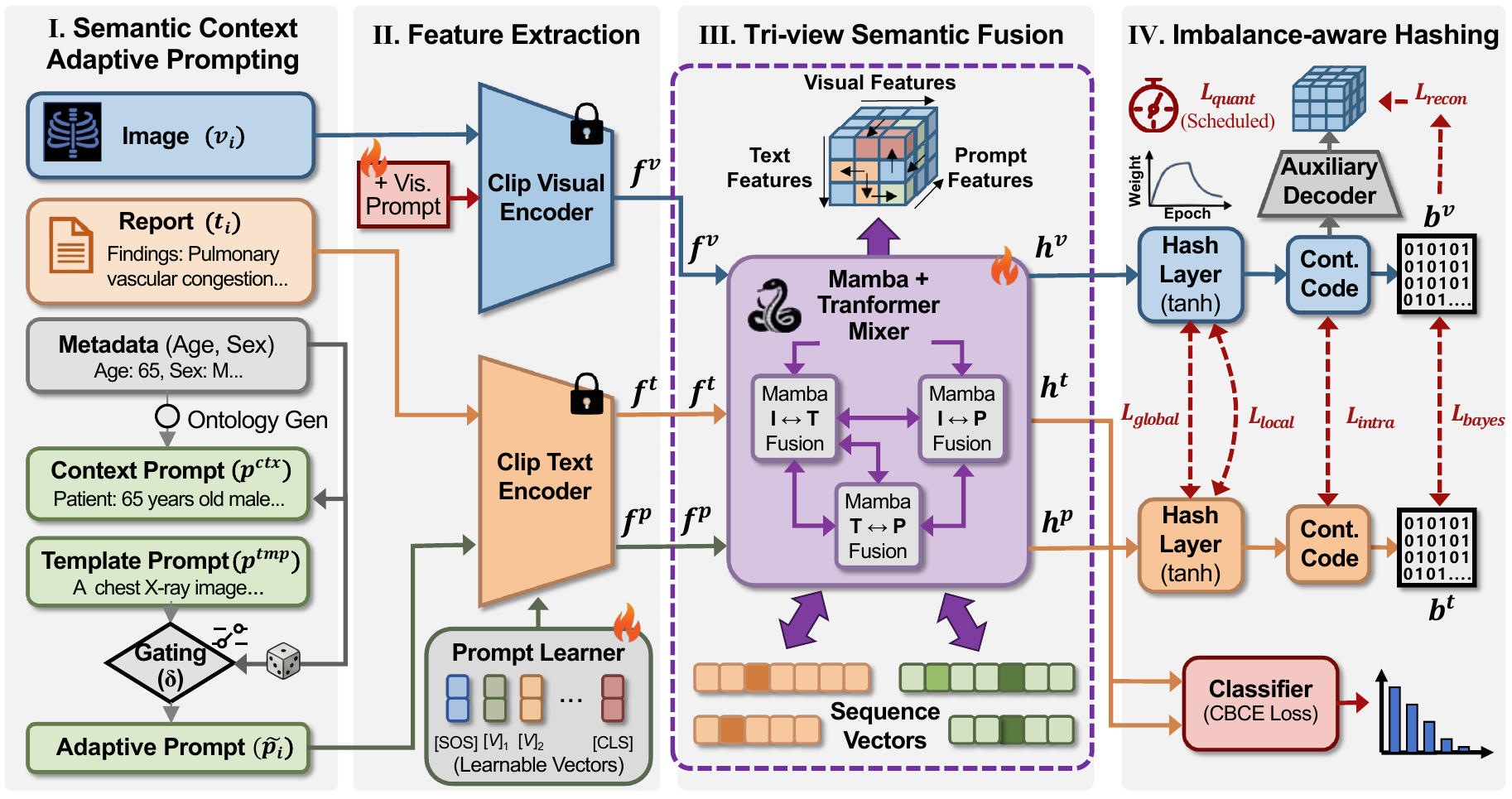} 
  \vspace{-0.2cm}
  \caption{The overall framework of TriPAH. It consists of four main components: 
 (I) Semantic Context Adaptive Prompting, where patient-level prompts $\tilde{p}_i$ are synthesized via stochastic gating to complement raw reports; 
 (II) Feature Extraction, utilizing CLIP encoders with visual prompt injection; 
 (III) Tri-view Semantic Fusion, employing Mamba-Transformer blocks for pairwise interaction among image text and prompt; and 
 (IV) Imbalance-aware Multi-task Hashing, where binary codes are optimized via asymmetric quantization and class-balanced supervision $\mathcal{L}_{cbce}$ to handle long-tailed medical distributions.}
  \vspace{-0.5cm}
  \label{fig:framework}
\end{figure*}

As illustrated in \textbf{Fig.~\ref{fig:framework}}, TriPAH consists of four synergistic components: (1) Semantic Context Adaptive Prompting, which synthesizes ontology-grounded patient prompts to mitigate textual noise; (2) Feature Extraction, utilizing CLIP encoders with visual prompt injection to derive robust embeddings; (3) Tri-view Semantic Fusion, a lightweight mixer coupling Image, Text, and Prompt views via Mamba-Transformer blocks; and (4) Imbalance-aware Multi-task Hashing, which optimizes binary codes using asymmetric quantization and rare-class supervision.

\subsection{Semantic Context Adaptive Prompting}
Standard cross-modal hashing methods often utilize static templates (e.g., ``A photo of [CLS]''), which fails to capture the intricate patient-level context essential for diagnosis. To bridge the semantic gap between visual phenotypes and textual descriptions, we introduce a patient-centric prompt mechanism.

\textbf{Ontology-grounded Prompt Synthesis.}
We construct a dynamic prompt set $\mathcal{P}$ by integrating ontology-grounded attributes (e.g., age, sex, laterality) extracted from metadata. Let $T^{raw}$ be the raw report and $T^{meta}$ be the structured metadata. We define the prompt function $\phi(\cdot)$ that maps metadata to natural language descriptors:
\begin{equation}
    p_i^{ctx} = \phi(T_i^{meta}) \oplus \text{``Findings:''} \oplus \text{Extract}(T_i^{raw}),
\end{equation}
where $\oplus$ denotes concatenation and $\text{Extract}(\cdot)$ filters key diagnostic terms.

\textbf{Stochastic Gating Mechanism.}
To prevent the model from overfitting to specific prompt patterns and to enhance robustness against missing metadata, we employ a stochastic gating strategy during training. We define a Bernoulli random variable $\delta \sim \text{Bernoulli}(\rho)$, where $\rho$ is the gating probability. The final prompt input $\tilde{p}_i$ is formulated as:
\begin{equation}
    \tilde{p}_i = \delta \cdot p_i^{ctx} + (1 - \delta) \cdot p^{tmp},
\end{equation}
where $p^{tmp}$ is a generic fallback prompt. This ensures the hashing network remains resilient to variable clinical contexts. The text $t_i$ and prompt $\tilde{p}_i$ are processed by a shared text encoder to yield features $\mathbf{f}_i^t \in \mathbb{R}^{D}$ and $\mathbf{f}_i^p \in \mathbb{R}^{D}$. Simultaneously, the image $v_i$ is processed by a visual encoder with learnable visual prompts injected into the ViT layers, yielding $\mathbf{f}_i^v \in \mathbb{R}^{D}$.
\subsection{Synergistic Tri-View Feature Amalgamation}
To overcome semantic fragmentation, purely contrastive alignment is insufficient. We propose a Tri-view Semantic Fusion module that treats the Image ($\mathbf{f}^v$), Text ($\mathbf{f}^t$), and Prompt ($\mathbf{f}^p$) as three complementary views. We employ a hybrid architecture combining State Space Models (SSM) for efficient long-sequence modeling and Transformers for precise global attention.

\textbf{Pairwise View Interaction.}
We construct a fusion graph where features interact in triplets: Image-Text ($v \leftrightarrow t$), Image-Prompt ($v \leftrightarrow p$), and Text-Prompt ($t \leftrightarrow p$). For any pair of modalities $A$ and $B$ (e.g., $v$ and $p$), we define a fusion block $\Psi(A, B)$.
Inspired by recent advances in selective state spaces \cite{Zou2025CVPR_PromptHash}, we propose a synergistic fusion block that couples selective sequence modeling with cross-modal interaction. Formally, let $\Phi_{\text{ssm}}(\cdot)$ denote the Mamba-based selective scanning transformation. The tri-view semantic amalgamation is defined as a composite function:

\begin{equation}
    \mathbf{Z}_{fuse} = \text{Attn}\Big( \underbrace{\Phi_{\text{ssm}}(\mathbf{Z}_A) + \mathbf{Z}_A}_{\text{Denoised Anchor}}, \mathbf{Z}_B \Big) + \big(\Phi_{\text{ssm}}(\mathbf{Z}_A) + \mathbf{Z}_A \big),
\end{equation}
where $\text{Attn}(\mathbf{Q}, \mathbf{K})$ utilizes the SSM-refined features as queries to retrieve complementary cues from the key/value modality $\mathbf{Z}_B$. This nested residual design ensures that the fusion is anchored on robust, noise-filtered representations.

\textbf{Residual Gating and Aggregation.}
To prevent the prompt view from dominating or introducing hallucinations, we apply a Global Response Normalization (GRN) based gating mechanism. The refined features for the three views are aggregated via a weighted residual connection:
\begin{equation}
    \mathbf{h}_i^{view} = \mathbf{f}_i^{view} + \gamma \cdot \sum_{m \in \mathcal{M} \setminus view} \text{GRN}(\Psi(\mathbf{f}_i^{view}, \mathbf{f}_i^{m})),
\end{equation}
where $view \in \{v, t, p\}$ and $\gamma$ is a learnable scale factor. This process yields three semantically aligned feature vectors: $\mathbf{h}_i^v$, $\mathbf{h}_i^t$, and $\mathbf{h}_i^p$.

\textbf{Hierarchical Alignment Objectives.}
We enforce alignment across these fused views using a multi-granularity contrastive objective. The global alignment loss $\mathcal{L}_{global}$ aligns the primary image and text representations using InfoNCE:
\begin{equation}
    \mathcal{L}_{global} = - \log \frac{\exp(\text{sim}(\mathbf{h}_i^v, \mathbf{h}_i^t) / \tau)}{\sum_{j=1}^N \exp(\text{sim}(\mathbf{h}_i^v, \mathbf{h}_j^t) / \tau)}.
\end{equation}
Complementing this, we introduce a local prompt alignment loss $\mathcal{L}_{local}$ to align the image with the patient-context prompt. First, we compute an adaptive temperature $\tau_{adp}$ utilizing Jensen-Shannon divergence (JSD) to account for prompt uncertainty:
\begin{equation}
    \tau_{adp} = \tau \times (1 + \text{JSD}(\mathbf{h}_i^v || \mathbf{h}_i^p))^{-1}.
\end{equation}
Then, the alignment loss is formulated to handle noisy correspondence by relaxing the constraint on uncertain triplets:
\begin{equation}
    \mathcal{L}_{local} = - \log \frac{\exp(\text{sim}(\mathbf{h}_i^v, \mathbf{h}_i^p) / \tau_{adp})}{\sum_{j=1}^N \exp(\text{sim}(\mathbf{h}_i^v, \mathbf{h}_j^p) / \tau_{adp})}.
\end{equation}
This ensures that images with high uncertainty in their visual features rely more loosely on the prompt alignment, preventing confident misalignment.

\begin{table*}[t]
\centering
\caption{Performance comparison (mAP $\uparrow$) of different cross-modal hashing methods on ODIR-5K\cite{odir2019}, MIMIC-CXR\cite{Johnson2019SciData_MIMICCXR}, and IU-Xray\cite{demner2015preparing} datasets. The best results are highlighted in \textbf{bold}, and the second-best results are \underline{underlined}.}
\label{tab:main_results}
\resizebox{\textwidth}{!}{%
\begin{tabular}{llc cccc cccc cccc}
\toprule
\multirow{2}{*}[-0.8ex]{\textbf{Task}} & \multirow{2}{*}[-0.6ex]{\textbf{Method}} & \multirow{2}{*}[-0.6ex]{\textbf{Reference}} & \multicolumn{4}{c}{\textbf{ODIR-5K}} & \multicolumn{4}{c}{\textbf{MIMIC-CXR}} & \multicolumn{4}{c}{\textbf{IU-Xray}} \\
\cmidrule(lr){4-7} \cmidrule(lr){8-11} \cmidrule(lr){12-15}
 &  &  & 32 bits & 64 bits & 128 bits & Mean & 32 bits & 64 bits & 128 bits & Mean & 32 bits & 64 bits & 128 bits & Mean \\
\midrule
\multirow{8}{*}{$I \to T$} 
 & DCHMT & MM'22 & 0.502 & 0.518 & 0.522 & 0.514 & 0.681 & \underline{0.693} & 0.697 & 0.690 & 0.741 & 0.746 & \underline{0.761} & 0.749 \\
 & MITH & MM'23 & 0.501 & 0.506 & 0.524 & 0.510 & 0.667 & 0.669 & 0.668 & 0.668 & 0.721 & 0.735 & 0.740 & 0.732 \\
 & DSPH & TCSVT'24 & 0.449 & 0.487 & 0.488 & 0.475 & 0.637 & 0.671 & 0.682 & 0.663 & 0.719 & 0.721 & 0.723 & 0.721 \\
 & CMCL & TKDE'24 & 0.531 & 0.527 & 0.542 & 0.533 & 0.668 & 0.672 & 0.669 & 0.670 & 0.713 & 0.715 & 0.717 & 0.715 \\
 & CICH & TKDE'24 & 0.542 & 0.528 & 0.525 & 0.532 & \underline{0.690} & \underline{0.693} & \underline{0.699} & \underline{0.694} & \underline{0.745} & \underline{0.753} & 0.754 & \underline{0.751} \\
 & MSACH & EAAI'25 & \underline{0.665} & \underline{0.669} & \underline{0.690} & \underline{0.675} & 0.686 & 0.689 & 0.699 & 0.691 & -- & -- & -- & -- \\
 & \textbf{TriPAH (Ours)} & \textbf{--} & \textbf{0.936} & \textbf{0.939} & \textbf{0.937} & \textbf{0.937} & \textbf{0.808} & \textbf{0.823} & \textbf{0.830} & \textbf{0.820} & \textbf{0.894} & \textbf{0.895} & \textbf{0.899} & \textbf{0.896} \\
 \cmidrule(lr){2-15}
 & \textit{Improvement} & -- & \textcolor{red}{+0.271} & \textcolor{red}{+0.270} & \textcolor{red}{+0.247} & \textcolor{red}{+0.262} & \textcolor{red}{+0.118} & \textcolor{red}{+0.130} & \textcolor{red}{+0.131} & \textcolor{red}{+0.126} & \textcolor{red}{+0.149} & \textcolor{red}{+0.142} & \textcolor{red}{+0.138} & \textcolor{red}{+0.145} \\
\midrule
\multirow{8}{*}{$T \to I$} 
 & DCHMT & MM'22 & 0.707 & 0.711 & 0.718 & 0.712 & 0.713 & 0.720 & 0.723 & 0.719 & \underline{0.836} & \underline{0.836} & \underline{0.836} & \underline{0.836} \\
 & MITH & MM'23 & 0.432 & 0.443 & 0.436 & 0.437 & 0.616 & 0.620 & 0.621 & 0.619 & 0.686 & 0.685 & 0.683 & 0.685 \\
 & DSPH & TCSVT'24 & 0.351 & 0.385 & 0.401 & 0.379 & 0.591 & 0.620 & 0.629 & 0.613 & 0.637 & 0.673 & 0.675 & 0.662 \\
 & CMCL & TKDE'24 & 0.390 & 0.393 & 0.407 & 0.397 & 0.624 & 0.632 & 0.627 & 0.628 & 0.660 & 0.655 & 0.655 & 0.657 \\
 & CICH & TKDE'24 & 0.813 & 0.819 & \underline{0.931} & 0.854 & \underline{0.841} & \underline{0.846} & \underline{0.876} & \underline{0.854} & 0.800 & 0.778 & 0.759 & 0.779 \\
 & MSACH & EAAI'25 & \underline{0.888} & \underline{0.889} & 0.901 & \underline{0.893} & 0.749 & 0.756 & 0.763 & 0.756 & -- & -- & -- & -- \\
 & \textbf{TriPAH (Ours)} & \textbf{--} & \textbf{0.989} & \textbf{0.971} & \textbf{0.972} & \textbf{0.977} & \textbf{0.868} & \textbf{0.898} & \textbf{0.901} & \textbf{0.889} & \textbf{0.892} & \textbf{0.899} & \textbf{0.896} & \textbf{0.896} \\
 \cmidrule(lr){2-15}
 & \textit{Improvement} & -- & \textcolor{red}{+0.101} & \textcolor{red}{+0.082} & \textcolor{red}{+0.041} & \textcolor{red}{+0.084} & \textcolor{red}{+0.027} & \textcolor{red}{+0.052} & \textcolor{red}{+0.025} & \textcolor{red}{+0.035} & \textcolor{red}{+0.056} & \textcolor{red}{+0.063} & \textcolor{red}{+0.060} & \textcolor{red}{+0.060} \\
\bottomrule
\end{tabular}%
}
  \vspace{-0.5cm}
\end{table*}

\textbf{Intra-modal Semantic Consistency.}
Although both report and prompt belong to the textual modality, they represent different granularities (instance-specific vs. ontology-general). To enforce consistency within the language branch, we minimize the distance between their fused representations:
\begin{equation}
    \mathcal{L}_{intra} = 1 - \frac{1}{N} \sum_{i=1}^N \frac{(\mathbf{h}_i^t)^\top \mathbf{h}_i^p}{\|\mathbf{h}_i^t\| \|\mathbf{h}_i^p\|}.
\end{equation}

\subsection{Imbalance-aware Multi-task Hashing}
The final stage projects the fused features into binary codes. Medical datasets are inherently long-tailed; standard hashing losses often bias towards majority classes, leading to the collapse of rare disease codes. We propose an imbalance-aware optimization strategy.

\textbf{Asymmetric Hash Projection.}
The continuous embeddings are projected to $K$ bits via a hashing layer: $\mathbf{z}_i^* = \text{tanh}(\mathbf{W}_h \mathbf{h}_i^* + \mathbf{b}_h)$, where $* \in \{v, t\}$. To obtain binary codes, we apply the sign function $\mathbf{b}_i = \text{sgn}(\mathbf{z}_i)$. However, direct optimization is intractable. We employ a relaxation strategy with an asymmetric quantization schedule.
For the text branch, which is semantically more stable, we apply a constant quantization constraint. For the image branch, we use a progressive annealing strategy:
\begin{equation}
    \mathcal{L}_{quant} = \sum_{i=1}^N \left( \|\mathbf{b}_i^t - \mathbf{z}_i^t\|^2 + \lambda(e) \|\mathbf{b}_i^v - \mathbf{z}_i^v\|^2 \right),
\end{equation}
where $\lambda(e)$ is a time-dependent weight that increases logistically over epochs $e$. This delays strong binary constraints on the image branch, preventing early discretization collapse.

\textbf{Class-Balanced Classification Supervision (CBCE).}
To explicitly enforce separability for rare diseases, we introduce a multi-label classification head on top of the hash codes. We utilize a Class-Balanced Cross-Entropy (CBCE) loss. Let $N_c$ be the number of samples for class $c$. The weighting factor is defined as $\omega_c = (1 - \zeta)/(1 - \zeta^{N_c})$, where $\zeta$ is a hyperparameter for class balance. The loss is:
\begin{equation}
    \mathcal{L}_{cbce} = - \sum_{i=1}^N \sum_{c=1}^C \omega_c \left[ y_{i,c} \log(\hat{y}_{i,c}) + (1-y_{i,c}) \log(1-\hat{y}_{i,c}) \right],
\end{equation}
where $\hat{y}_{i}$ is the prediction derived from the hash codes. This forces the hash bits to encode discriminative features even for under-represented classes.

\textbf{Bayesian Consistency Regularization.}
Finally, to maintain structure in the Hamming space, we apply a Bayesian consistency loss minimizing the Kullback-Leibler (KL) divergence between the inter-modal code distribution $P(\mathbf{b}^v | \mathbf{b}^t)$ and the prior label similarity distribution $S_{ij}$:
\begin{equation}
    \mathcal{L}_{bayes} = \sum_{i,j} \text{KL}(S_{ij} || \sigma(\frac{1}{K} (\mathbf{b}_i^v)^\top \mathbf{b}_j^t)),
\end{equation}
where $\sigma$ is the sigmoid function.

\subsection{Total Optimization Objective}
The overall training objective of TriPAH integrates the alignment, hashing, and classification tasks:
\begin{equation}
\begin{split}
    \mathcal{L}_{total} &= \mathcal{L}_{global} + \alpha \mathcal{L}_{local} + \mu \mathcal{L}_{intra} \\
    &\quad + \kappa \mathcal{L}_{quant} + \xi \mathcal{L}_{cbce} + \eta \mathcal{L}_{bayes},
\end{split}
\end{equation}
where $\alpha, \mu, \kappa, \xi, \eta$ are hyperparameters balancing the contributions of each component. By jointly optimizing these objectives, TriPAH learns binary codes that are semantically aligned across three views and robust to clinical noise and data imbalance.

\begin{figure*}[t!]
  \centering
  \includegraphics[width=1.0\textwidth]{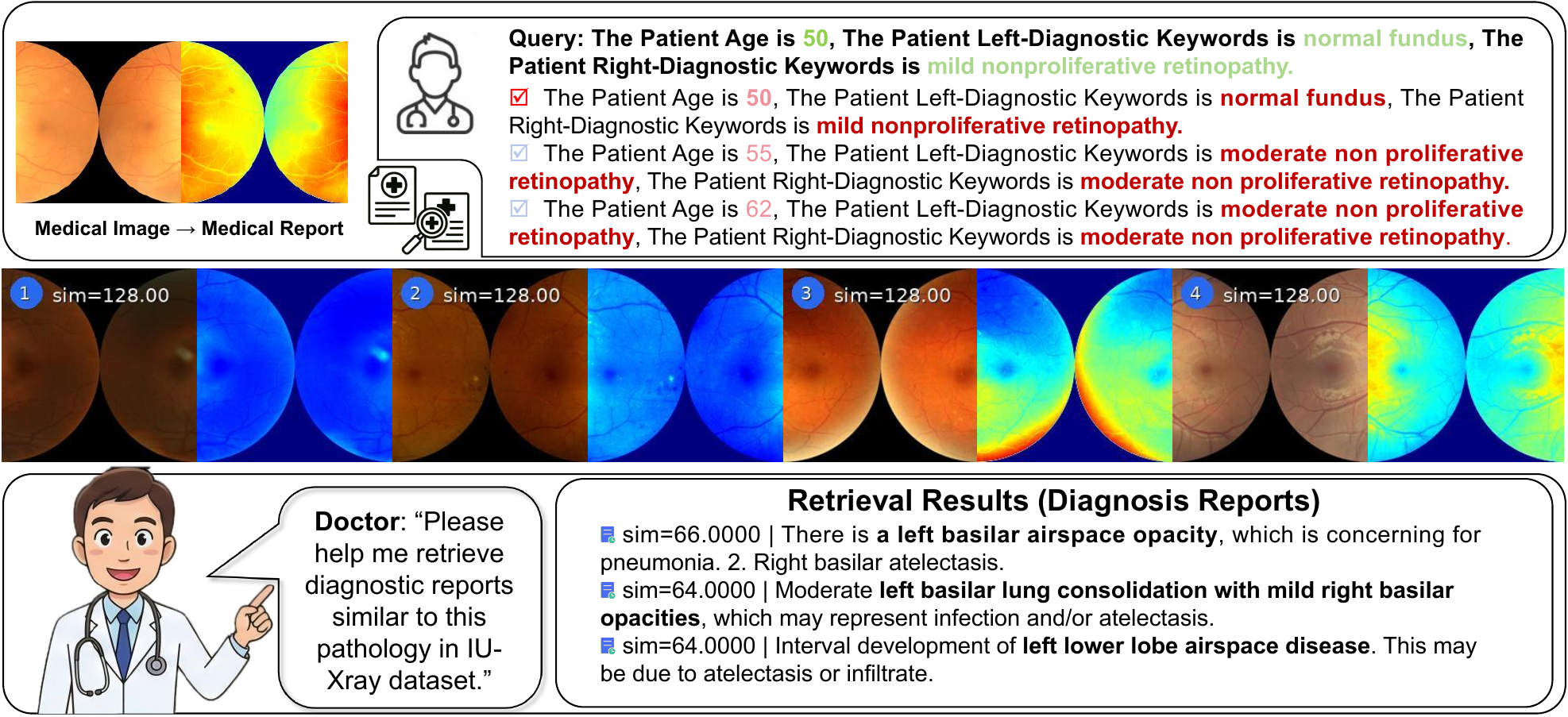}
   \vspace{-0.2cm}
  \caption{Qualitative visualization of cross-modal retrieval results on heterogeneous medical datasets. \textbf{(Top)} On the ODIR-5K dataset (fundus images), TriPAH accurately retrieves reports matching specific diagnostic keywords like ``normal fundus'' and ``retinopathy'' based on visual cues. \textbf{(Bottom)} On the IU-Xray dataset (chest X-rays).}
   \vspace{-0.5cm}
  \label{fig:demo_results}
\end{figure*}
\section{Experiments}

\subsection{Datasets and Implementation Details}

We evaluate TriPAH on three public medical datasets: \textbf{ODIR-5K} (Ophthalmic), \textbf{MIMIC-CXR} (Chest X-ray), and \textbf{IU-Xray} (Chest X-ray).
Statistics for the training, query, and retrieval splits are detailed in Table~\ref{tab:datasets}.
ODIR-5K contains binocular fundus images with 8 disease labels.
MIMIC-CXR is a large-scale dataset with multi-label chest radiographs, exhibiting a severe long-tailed distribution.
IU-Xray is a smaller dataset but includes rich textual reports.
For ODIR-5K, we synthesized patient-level prompts using metadata (age, sex, diagnostic keywords).
For chest X-ray datasets, we utilized the ``Findings'' and ``Impression'' sections as textual input.

\begin{center} 
\captionof{table}{Summary of Statistics for Three Benchmark Datasets.} 
\label{tab:datasets}
\begin{tabular}{lcccc}
\toprule
\textbf{Dataset} & \textbf{Training} & \textbf{Query} & \textbf{Retrieval} & \textbf{Total} \\
\midrule
IU X-Ray   & 2,500   & 380    & 3,446   & 3,826   \\
ODIR-5K    & 2,275   & 350    & 3,150   & 3,500   \\
MIMIC-CXR  & 100,000 & 20,000 & 207,814 & 227,814 \\
\bottomrule
\end{tabular}
\end{center}

We implement TriPAH in PyTorch on an NVIDIA RTX 5090 GPU using a pre-trained CLIP (ViT-B/16) backbone. Images are resized to $224 \times 224$. The model is trained for 100 epochs using AdamW optimizer (batch size 128) with learning rates of $1 \times 10^{-5}$ for the backbone and $1 \times 10^{-4}$ for hashing heads. Hyperparameters are set to $\tau=0.05$, $\alpha=1.0$, $\mu=1.0$, $\kappa=0.1$ (dynamic), $\xi=0.3$, and $\eta=0.01$.

\subsection{Comparison with State-of-the-Art}

We compare TriPAH with six recent state-of-the-art cross-modal hashing baselines, including DCHMT \cite{Tu2022ACMMM_DCHMT}, MITH \cite{Liu2023ACMMM_MITH}, DSPH \cite{Huo2024TCSVT_DSPH}, CMCL \cite{Wu2024TKDE_CMCL}, CICH \cite{Luo2024TKDE_CICH}, and the medical baseline MSACH \cite{Liu2025EAAI_MSACH}.
Table~\ref{tab:main_results} reports the mAP scores for both Image-to-Text ($I \to T$) and Text-to-Image ($T \to I$) retrieval tasks at 32, 64, and 128 bits.

\noindent\textbf{Performance Analysis.}
TriPAH consistently outperforms all baselines across all datasets and code lengths.
On the \textbf{ODIR-5K} dataset, our method achieves a remarkable improvement, surpassing the second-best method (MSACH) by an average of \textbf{+26.2\%} in $I \to T$ mAP.
This demonstrates that our Tri-view Fusion module effectively captures fine-grained lesions in fundus images that are often missed by standard dual-stream methods.
On the large-scale \textbf{MIMIC-CXR} dataset, TriPAH achieves a mean mAP of \textbf{0.820} ($I \to T$) and \textbf{0.889} ($T \to I$), outperforming CICH by significant margins.
This indicates that our imbalance-aware optimization strategy (CBCE) successfully mitigates the bias towards majority classes in long-tailed medical data.
Furthermore, TriPAH maintains high stability across different bit lengths, suggesting that the generated binary codes are compact and semantically rich.

\subsection{Efficiency Analysis}
\begin{table}[t]
\centering
\caption{Efficiency comparison in terms of training and inference time.}
\label{tab:time}
\resizebox{0.48\textwidth}{!}{
\begin{tabular}{l|cc|cc|cc}
\toprule
 & \multicolumn{2}{c|}{ODIR} & \multicolumn{2}{c|}{IU-Xray} & \multicolumn{2}{c}{MIMIC-CXR} \\
Method & Train & Infer. & Train & Infer. & Train & Infer. \\
\midrule
DCHMT & 3200s & 111s & 2550s & 94s & 13950s & 492s \\
DSPH  & 2800s & 106s & 2500s & 92s & 10650s & 1014s \\
MITH  & 1950s & 80s  & 2550s & 73s & 16550s & \textbf{180s} \\
CMCL  & 1700s & 54s  & 1300s & 34s & 32200s & 953s \\
\midrule
\rowcolor[HTML]{FFE6EF}
\textbf{TriPAH (Ours)} & \textcolor{red}{\textbf{516s}} & \textcolor{red}{\textbf{19s}} & \textcolor{red}{\textbf{202s}} & \textcolor{red}{\textbf{8s}} & \textcolor{red}{\textbf{4785s}} & 329s \\
\bottomrule
\end{tabular}
}
\vspace{-2.2em}
\end{table}
We further evaluate the computational efficiency of TriPAH in terms of training and inference time. All efficiency results are measured on the same hardware platform under identical data loading settings. The reported inference time includes forward code generation and database retrieval, reflecting end-to-end practical deployment cost rather than isolated backbone latency alone. 

As shown in Table~\ref{tab:time}, TriPAH achieves the fastest training and inference on ODIR and IU-Xray, and substantially reduces training time on MIMIC-CXR while maintaining competitive inference efficiency. This gain mainly comes from the hybrid design, where Mamba-based selective sequence modeling scales linearly with sequence length and avoids heavy full self-attention. These results show that TriPAH is not only accurate but also computationally efficient in practice.

\subsection{Ablation Studies}

To validate the contribution of each component, we conduct ablation studies on three variants: (A) Baseline (CLIP+Hash), (AC) Baseline + IA-Hash, and (AB) Baseline + Tri-view Fusion. The results are summarized in Table~\ref{tab:ablation_study}.

\noindent\textbf{Impact of Tri-view Semantic Fusion.}
Comparing Variant A and AB, incorporating the Tri-view Fusion module yields the most significant performance boost (e.g., +0.257 mean mAP on ODIR).
This confirms that introducing the prompt as a third ``anchor" view effectively bridges the semantic gap between visual features and noisy clinical reports, preventing semantic fragmentation.

\noindent\textbf{Impact of Imbalance-aware Hashing.}
Comparing Variant A and AC, adding the imbalance-aware objectives (CBCE + Quantization Schedule) provides consistent improvements (e.g., +0.027 mean mAP on ODIR).
More importantly, when combined with fusion (TriPAH vs. AB), IA-Hash further refines the retrieval performance (e.g., +0.022 mean mAP on MIMIC), particularly for identifying rare disease cases which are critical for clinical diagnosis.
The full model (TriPAH) achieves the best performance, validating the synergy between semantic alignment and robust hashing optimization.

\begin{table*}[t]
\centering
\caption{Ablation study of different components on ODIR-5K, MIMIC-CXR, and IU-Xray datasets. \textbf{Tri-view}: Tri-view Semantic Fusion; \textbf{IA-Hash}: Imbalance-aware Multi-task Hashing. The best results are highlighted in bold.}
\label{tab:ablation_study}
\resizebox{1.0\textwidth}{!}{
\begin{tabular}{l c c ccc ccc ccc}
\toprule
\multirow{2}{*}[-0.8ex]{\textbf{Variant}} 
& \multirow{2}{*}[-0.8ex]{\textbf{Tri-view}} 
& \multirow{2}{*}[-0.8ex]{\textbf{IA-Hash}} 
& \multicolumn{3}{c}{\textbf{ODIR-5K (mAP)}} 
& \multicolumn{3}{c}{\textbf{MIMIC-CXR (mAP)}} 
& \multicolumn{3}{c}{\textbf{IU-Xray (mAP)}} \\
\cmidrule(lr){4-6} \cmidrule(lr){7-9} \cmidrule(lr){10-12}
 & & & $I \to T$ & $T \to I$ & Mean 
 & $I \to T$ & $T \to I$ & Mean 
 & $I \to T$ & $T \to I$ & Mean \\
\midrule
A (Baseline) & $\times$ & $\times$ & 0.641 & 0.709 & 0.675 & 0.685 & 0.691 & 0.688 & 0.732 & 0.711 & 0.722 \\
AC (w/o Tri-view) & $\times$ & $\checkmark$ & 0.678 & 0.725 & 0.702 & 0.705 & 0.712 & 0.709 & 0.764 & 0.735 & 0.750 \\
AB (w/o IA-Hash) & $\checkmark$ & $\times$ & 0.911 & 0.952 & 0.932 & 0.801 & 0.865 & 0.833 & 0.877 & 0.875 & 0.876 \\
\textbf{TriPAH (ABC)} & $\checkmark$ & $\checkmark$ & \textbf{0.937} & \textbf{0.977} & \textbf{0.957} & \textbf{0.820} & \textbf{0.889} & \textbf{0.855} & \textbf{0.896} & \textbf{0.896} & \textbf{0.896} \\
\bottomrule
\end{tabular}%
}
\vspace{-0.5cm}
\end{table*}

\subsection{Qualitative Results}
Figure~\ref{fig:demo_results} visualizes top-3 retrieval results on ODIR-5K and IU-Xray.
As shown in the top row (ODIR-5K), given a fundus image with early-stage retinopathy, baseline methods tend to retrieve reports describing generic ``normal" eyes.
In contrast, TriPAH accurately retrieves reports containing specific keywords like ``mild nonproliferative retinopathy", matching the visual evidence.
Similarly, in the chest X-ray example (bottom row), our method correctly identifies ``airspace opacity" and ``pneumonia" despite the visual complexity.
These results intuitively verify that TriPAH learns semantically aligned and medically accurate binary codes.

\section{Conclusion}
We proposed TriPAH, an imbalance-aware framework designed to tackle semantic fragmentation and long-tailed distributions in medical retrieval. 
By synergizing stochastic patient-context prompts with Mamba-based tri-view fusion, TriPAH bridges the gap between visual phenotypes and noisy reports. 
Moreover, our optimization strategy, incorporating asymmetric quantization and class-balanced supervision, ensures discriminative codes for rare diseases. 
Experiments on ODIR-5K, MIMIC-CXR, and IU-Xray demonstrate that TriPAH significantly outperforms state-of-the-art baselines. 
Future work will explore integrating foundation models to enhance generalization.

\begingroup
\small
\renewcommand{\baselinestretch}{0.85}\selectfont
\let\OLDthebibliography\thebibliography
\renewcommand\thebibliography[1]{
  \OLDthebibliography{#1}
  \setlength{\parskip}{0pt}
  \setlength{\itemsep}{0pt plus 0.3ex}
}

\bibliographystyle{IEEEtran}
\bibliography{refs} 
\endgroup

\end{document}